\documentclass[report,twocolumn,amsmath,amsfonts,amssymb,showpacs,a4]{revtex4}
\usepackage{epsfig,psfrag,graphicx,bm,color,verbatim}

\def\msun{{\rm\,M_\odot}}

\begin{document}
\title{Dark Matter Annihilation around Intermediate Mass Black Holes: an update}

\author{Gianfranco Bertone $^a$, Mattia Fornasa$^b$, Marco Taoso$^b$ \& Andrew R. Zentner$^c$}
\affiliation{$^a$ Institut d'Astrophysique de Paris, UMR 7095-CNRS \\ 
Universit\'e Pierre et Marie Curie, Bd Arago 98bis, 75014 Paris \\ 
$^b$ University of Padova \& INFN sezione di Padova, via Marzolo 8, 35131 Padova, Italy \\
$^c$ Dept. of Physics and Astronomy, University of Pittsburgh, Pittsburgh, PA 15260 USA}

\begin{abstract}
The formation and evolution of Black Holes inevitably affects the distribution
of dark and baryonic matter in the neighborhood of the Black Hole. These 
effects may be particularly relevant around Supermassive and Intermediate Mass 
Black Holes (IMBHs), the formation of which can lead to large Dark Matter 
overdensities, called {\em spikes} and {\em mini-spikes} respectively.  
Despite being larger and more dense, spikes evolve at the very centers of 
galactic halos, in regions where numerous dynamical effects tend to destroy 
them. Mini-spikes may be more likely to survive, and they have been proposed 
as worthwhile targets for indirect Dark Matter searches. We review here the 
formation scenarios and the prospects for detection of mini-spikes, and we 
present new estimates for the abundances of mini-spikes to illustrate the 
sensitivity of such predictions to cosmological parameters and uncertainties 
regarding the astrophysics of Black Hole formation at high redshift.  We also 
connect the IMBHs scenario to the recent measurements of cosmic-ray electron 
and positron spectra by the PAMELA, ATIC, H.E.S.S., and Fermi collaborations.
\end{abstract}

\maketitle

\section{Introduction}
\label{sec:Introduction}

The distribution of matter around a Black Hole (BH) is inevitably affected
by its formation and evolution, both in the case of baryons
\cite{Peebles:1972,Young:1980} and Dark Matter (DM) \cite{Gondolo:1999ef}.
The physics is relatively simple: when a compact object forms, the 
surrounding matter reacts to the increased gravitational potential, and if 
the growth proceeds on a timescale much longer than the dynamical time of 
the system, large overdensities can be achieved. This is particularly 
important for the DM distribution, because the DM annihilation rate is 
proportional to the square of the number density of DM particles. This means 
that if BH formation increases the DM density locally, it also boosts the 
annihilation rate and improves the prospects for detecting DM through the 
observation of secondary particles such as gamma-rays, anti-matter and 
neutrinos. 

Gondolo and Silk have applied this argument to the distribution of DM at the 
center of the Galaxy \cite{Gondolo:1999ef}, where a Supermassive BH (SMBH) of 
2--4 $ \times 10^6 M_\odot $ is known to dominate the gravitational potential 
within a sphere of radius $ r_h \sim 1 $~pc. The consequent enhancement in the 
DM density around the central SMBH (dubbed a {\it spike}) was subsequently 
shown likely to be reduced by  major merger events, off-center formation of 
the seed BH, gravitational scattering off stars and DM annihilations
\cite{Ullio:2001fb,Merritt:2002vj,Bertone:2005hw,Bertone:2005xv}.  
Zhao and Silk have subsequently suggested that mild overdensities of DM could 
also be present around Intermediate Mass Black Holes (IMBHs), remnants of 
Pop III stars, and provided the first rough estimates of the properties of 
{\it mini-spikes} surrounding IMBHs in the Milky Way \cite{Zhao:2005zr}.

Bertone, Zentner and Silk (hereafter BZS) built a consistent formation
and evolution scenario for IMBHs and associated mini-spikes, for two different 
classes of models: ``mild'' mini-spikes around IMBHs remnants of Pop III 
stars, and ``strong'' mini-spikes, around more massive compact objects 
\cite{Bertone:2005xz}. For both scenarios, BZS tracked the merger history of 
each individual BH, and selected precisely those IMBHs which never 
experienced mergers, to ensure at a minimum that major mergers have not 
destroyed any spike that may have existed around the original BH. This has 
allowed a detailed quantitative description of the mini-spikes scenario, and 
practically all subsequent studies published in the literature have made use 
directly or indirectly of the BZS IMBH models. Moreover, BZS were able to 
predict some of the demographics of wandering IMBHs, that may not be directly 
associated with the Milky Way bulge or disk, and they proposed to search for 
a class of gamma-rays objects sharing the same energy spectrum as a 
smoking-gun signature of DM annihilations. 

Here, we review the formation scenarios and the physics behind the BZS IMBH 
models, and explore how the population of IMBHs, thus of mini-spikes, depends 
on cosmological, astrophysical and particle physics parameters. To this aim, 
we update the original BZS catalogs, including the most recent determination
of cosmological parameters, and explore the dependence on astrophysical 
quantities. 

We also review here the experimental strategies that may allow the detection 
of mini-spikes, including the observation of multiple gamma-ray and neutrino 
sources in the Milky Way, a collection of gamma-ray sources around M31, and 
the possible contribution to the absolute flux and the angular power spectrum 
of the Extra-galactic Gamma-ray Background (EGB).

The paper is organized as follows: in Section \ref{sec:Formation_scenarios},
we review the IMBHs formation scenarios and discuss the current population of 
IMBHs in the Milky Way. In Section \ref{sec:Adiabation_contraction} we 
discuss adiabatic contraction and the formation of mini-spikes. 
Section \ref{sec:Galactic_IMBHs} is dedicated to the prospects for detecting
DM through gamma-rays and neutrinos from IMBHs in our Galaxy and in M31. In 
Section \ref{sec:Angular_spectrum} we discuss the impact of the mini-spikes 
population on the Extra-galactic Gamma-ray Background. In the final section 
(Section \ref{sec:Discussion_conclusions}) we discuss our results and 
present our conclusions.

\section{Intermediate Mass Black Holes}
\label{sec:Formation_scenarios}

\begin{figure*}
\begin{center}
\includegraphics[width=0.45\textwidth]{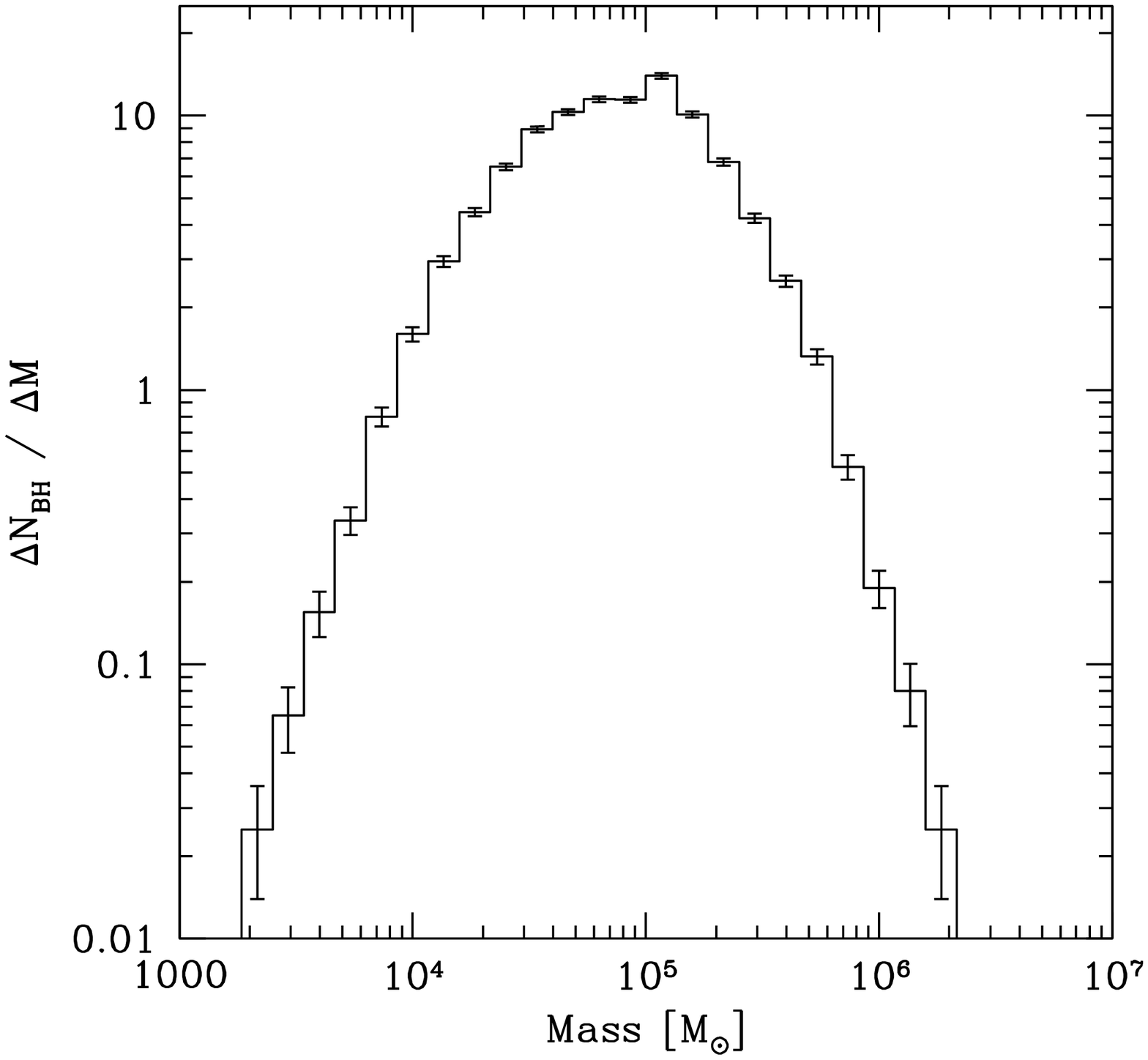}
\includegraphics[width=0.45\textwidth]{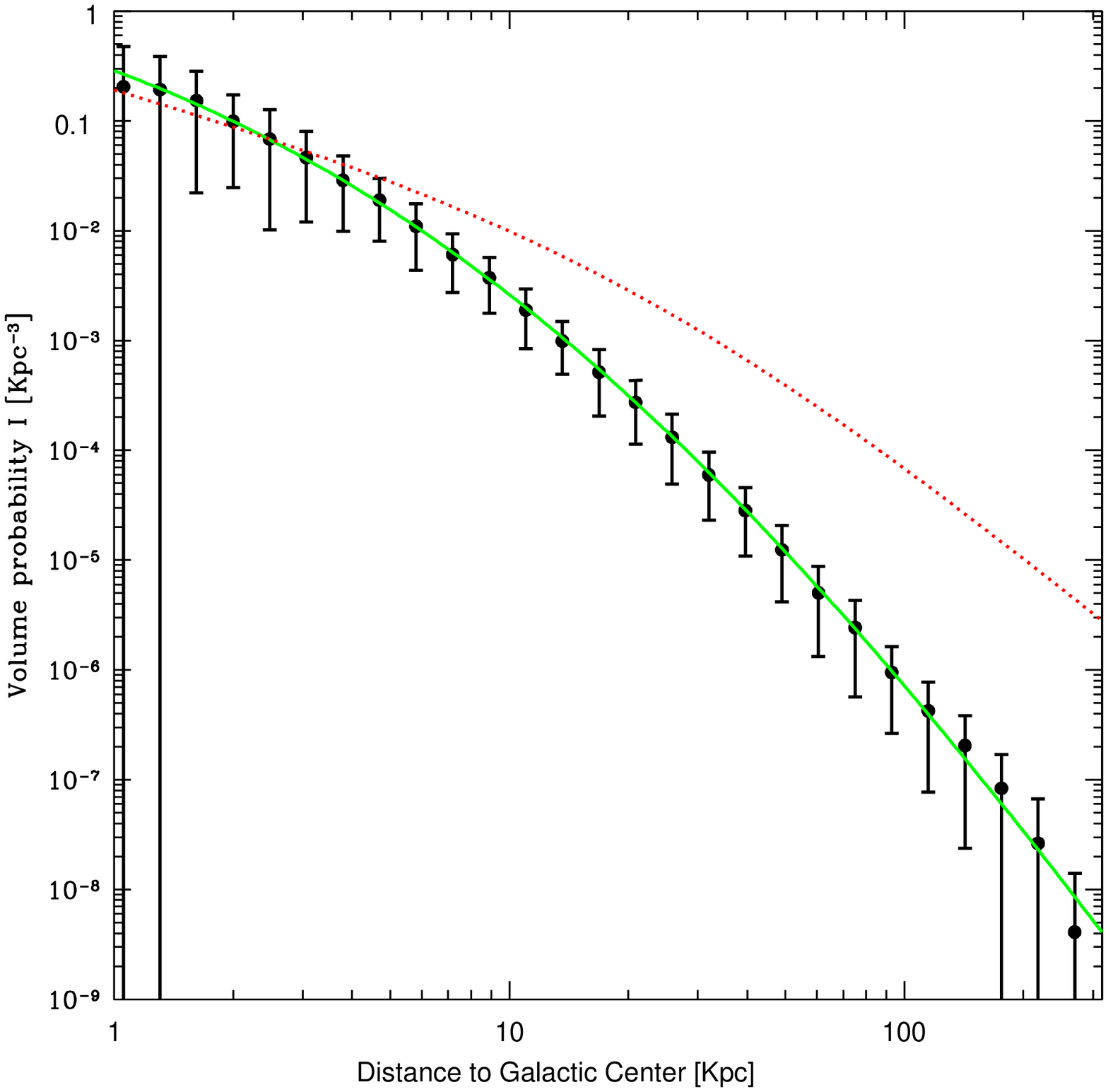}
\caption{\label{fig:mass_spectrum} {\it Left:} Mass function of unmerged IMBHs in the scenario B for a MW halo at $ z=0 $. The distribution is based on the average of 200 Monte Carlo realizations of a halo of virial mass $ M_{vir}=10^{12.1} h^{-1} M_\odot $, roughly the size of the halo of the MW. Taken from Ref. \cite{Bertone:2005xz}. {\it Right:} Radial distribution of the IMBH population in the MW from the numerical simulation of Ref. \cite{Bertone:2005xz}. The points refers to the average among 200 Monte Carlo realizations of the MW halo and the error bars show the scatter among realizations. The solid line is the analytical fit and the dotted line is a NFW profile. Taken from Ref. \cite{Taoso:2008qz}.}
\end{center}
\end{figure*}

\subsection{Hints for detection of IMBHs}
We refer to BHs with masses within a relatively large range, 
$ 100 M_\odot \lesssim M \lesssim 10^6 M_\odot $ as Intermediate Mass Black
Holes.  The lower limit is roughly the mass of the largest BH that can be 
produced by stellar collapse \cite{Fryer:1999ht}, and the upper limit 
corresponds to the minimum mass for a SMBH. There is not a clear experimental 
evidence for this class of objects, but they are often invoked as an 
explanation of Ultra Luminous X-ray sources, i.e. sources that emit at a 
luminosity larger than $ 10^{39} \mbox{erg s}^{-1} $, thus exceeding the 
Eddington luminosity of stellar BHs \cite{Swartz:2004xt}.

IMBHs may also be hosted inside globular clusters \cite{Miller:2003sc}. 
A hint in favor of this possibility comes from the fact that the predicted 
masses for the IMBHs and the velocity dispersion typical of a globular 
cluster follow exactly the extrapolation at lower values of the 
$ M_\bullet-\sigma $ relation valid for SMBHs \cite{Ferrarese:2004qr}.

From a theoretical point of view, IMBHs may provide massive enough seeds for 
the growth of high-redshift SMBHs, as observed by the Sloan Digital Sky Survey
in the form of quasars at redshifts as high as $ z \sim 6 $ 
\cite{Fan:2001ff,Bath:2003,Willott:2003xf}. Conclusive evidence for IMBHs may 
come in the near future from the detection (e.g. by LISA \cite{LISA}) of the 
gravitational waves emitted in the mergers of IMBHs 
\cite{Koushiappas:2005qz,Flanagan:1997sx,Flanagan:1997kp}.

\subsection{Formation}
In BZS, two formation scenarios for IMBHs have been studied, and we refer to 
that paper for a detailed discussion. In the first scenario (scenario A) 
IMBHs are remnants of the collapse of Population III stars 
\cite{Madau:2001sc}. Since the Jeans mass scales with the temperature as
$ T^{3/2} $, the mass of Pop III stars (which form early in a hot environment)
is expected to be higher than other stars, around $ 150-200 \mbox{ } M_\odot $ 
\cite{Abel:2000tu,Bromm:2001bi}. 
Newtonian simulations suggest that the evolution of Pop~III stars is very
different from stars in the local Universe. Zero metallicity Pop~III stars 
evolve on a timescale of order $ t_{*} \sim 1-10 $~Myr.  If their masses are 
in the range $ M \sim 60-140 M_\odot $ and $ M \gtrsim 260 M_\odot $, they 
collapse directly to BHs, while if 
$ 140 \lesssim M/M_\odot \lesssim 260 M_\odot $, they are completely disrupted 
due to the pulsation-pair-production instability, leaving behind no remnant
\cite{Heger:2002by} (see also 
Refs. \cite{Bond:1984,Fryer:2000my,Larson:1999zs,Schneider:1999us}).  

The second formation scenario (scenario B) considered by BZS was based on 
Ref.\cite{Koushiappas:2003zn}, as an example of models in which the collapse 
of primordial gas in early-forming halos leads to the ``direct'' formation of 
very massive objects 
\cite{Haehnelt:1993,Loeb:1994,Eisenstein:1995,Haehnelt:1998,Gnedin:2001ey,Bromm:2002hb}. 
In this specific scenario, the baryons in the low tail of the angular 
momentum distribution form a pressure-supported disc at the center of the
halo. Gravitational instabilities then turn on an effective viscosity that
make baryons lose their angular momentum, causing an inward mass flow. The
process comes to an end after $ 1-30 \mbox{ Myr} $, when the first stars form 
or when the halo experiences a major merger \cite{Koushiappas:2003zn}. The 
central object will then undergo gravitational collapse producing the final BH. 

The requirement that loss of angular momentum via viscosity is effective, 
fixes a mass scale of $ 10^7-10^8 M_\odot $ for the initial halos, and leads 
to a mass scale for the final IMBHs of order $ 10^5 M_\odot $. In fact, the 
mass distribution for scenario B turns out to be a log-normal Gaussian 
($ \sigma_\bullet=0.9 $) with a mean value $ M_\bullet $:
\begin{eqnarray}
\label{massfunc}
M_\bullet & = & 3.8 \times 10^4 M_\bullet \left( \frac{\kappa}{0.5} \right)
\left( \frac{f}{0.03} \right)^{3/2} \left( \frac{M_{vir}}{10^7 M_\odot}
\right) \\
& & \left( \frac{1+z_f}{18} \right) \left( \frac{t}{10 \mbox{ Myr}} \right) \, ,
\nonumber
\end{eqnarray}
where $ M_{vir} $ is the virial mass of the halo, $ z_f $ the redshift of 
formation, $ f $ the fraction of the baryonic mass of the halo that goes into 
the central pressure-supported disc, $ \kappa $ the fraction of the mass of 
the disc that ends up in the final IMBH and $ t $ is the timescale available 
for angular momentum transfer due to the effective viscosity.
\begin{figure*}[t]
\begin{center}
\includegraphics[width=0.5\textwidth]{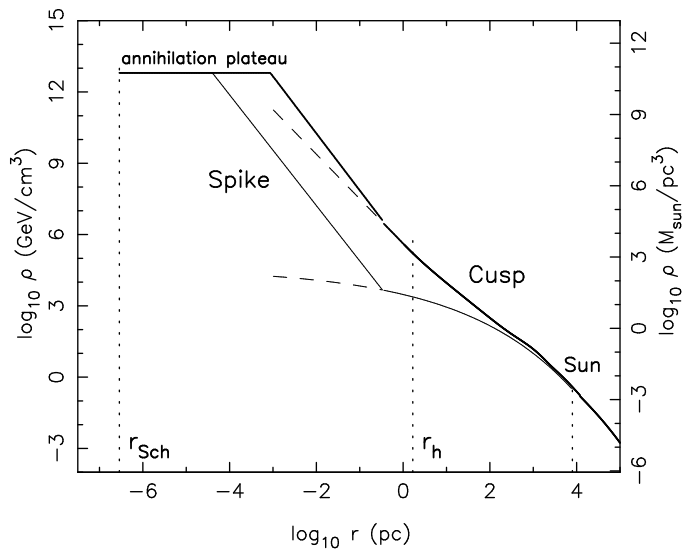}
\caption{\label{fig:spike} Possible models for the DM distribution in the Galaxy. The thin curve shows the standard halo model, and the thick curve is the same model after ``adiabatic compression'' by the Galactic baryons (stars and gas). Both curves are normalized to a DM density of $ 0.3 $ GeV cm$ ^{-3} $ at the Solar circle. Curves labeled ``spike'' show the increase in density that would result from growth of a Galactic SMBH at a fixed location. The annihilation plateau, $ \rho=\rho_a=m/\langle\sigma v\rangle t $, was computed assuming $ m=200 $ GeV, $ \langle\sigma v\rangle=10^{-28} $ cm$ ^3 $ s$ ^{-1} $ and $ t=10^{10} $ yr. Dotted vertical lines indicate the SMBH Schwarzschild radius ($ r_{Sch}\approx 2.9\times 10^{-7} $ pc, assuming a mass of $ 3.0\times 10^6M_\odot $ and zero rotation), the SMBH gravitational influence radius ($ r_h\approx 1.7 $ pc), and the radius of the Solar circle ($ R_\odot\approx 8.0 $ kpc). Effects of the dynamical processes (scattering of DM off stars, loss of DM into the SMBH, etc.) are excluded from this plot; these processes would generally act to decrease the DM density below what is shown here, particularly in the models with a ``spike.''. Taken from \cite{Bertone:2005xv}.}
\end{center}
\end{figure*}

BZS have estimated the population of IMBHs in the MW in scenario A as follows: 
they populated all the halos at $ z=18 $ that constitute a $ 3\sigma $ peak 
(with respect to the average density) with a pristine IMBH of mass 
$ 100 \mbox{ } M_\odot $. The evolution of the halos is then simulated as in 
Refs. \cite{Koushiappas:2005qz,Zentner:2003yd,Zentner:2004dq,Koushiappas:2003bn},
a procedure that tracks the growth and mergers of the structures down to 
redshift zero. The IMBHs in the halos that never experience mergers survive 
until the present epoch and constitute the current population of scenario A 
IMBHs. Their mass is assumed to remain unchanged with respect to the initial 
value $ \sim 100 \mbox{ } M_\odot $, and the number of unmerged IMBHs was 
found to be $ N_A = (1027 \pm 84) $. 

In the case of scenario B, BZS have followed a similar approach for the 
evolution of IMBHs, with the difference that only those halos with a mass
larger than the mass threshold $ M_{thr} $ for the onset of the effective 
viscosity (see the discussion above) are populated with IMBHs, with a mass 
function as in Eq. \ref{massfunc} (see also Fig. \ref{fig:mass_spectrum}). 
The predicted number of IMBHs in the MW for scenario B was 
$ N_B = (101 \pm 22) $ \cite{Bertone:2005xz} (see below for more recent 
results obtained with updated cosmological parameters). 

Each of these models of high-redshift seed BH formation is reasonable, 
but both of these scenarios remain uncertain in their detail. In particular, 
these models exhibit parameters that may reasonably take on a wide range of 
values leading to markedly different predictions for IMBH demographics at 
$ z=0 $.  The formation of seed BHs in both scenarios is cut off at redshifts 
when significant small-scale fragmentation of baryonic disks sets in and 
BH formation in scenario B terminates absolutely when the Universe becomes 
significantly reionized.  In what follows we exhibit the significant 
parameter freedom within these models by varying the formation redshift in 
scenario A, and the reionization redshift in scenario B among viable and 
illustrative values.  In scenario A, lowering the formation redshift requires 
that the seed black holes generally form in halos of higher mass, leading 
to reduced abundance, while in scenario B lowering the redshift of reionization 
allows angular momentum loss to be effective for a longer period, increasing 
the number of seed black holes.  We note here that the range of parameters 
we choose are currently viable and refer the reader to BZS for modeling details.

The radial distribution of IMBHs was found to be steeper than a 
Navarro-Frenk-White (NFW) profile \cite{Navarro:1996gj} (see the right panel 
of Fig.~\ref{fig:mass_spectrum}), and it is well fitted by the following 
analytical function \cite{Taoso:2008qz}:
\begin{equation}
N(r)=5.96 \times 10^{-2} \left[ 1+\Bigg(\frac{r}{9.1 \mbox{ kpc}}\Bigg)^{0.51} \right]^{-10.8}
\mbox{kpc}^{-3}.
\end{equation}

\section{Adiabatic contraction}
\label{sec:Adiabation_contraction}
As mentioned in the introduction, IMBHs and SMBHs are particularly interesting 
for indirect DM searches, because they may act as \emph{annihilation boosters}, 
as their growth can lead to large DM overdensities that enhance annihilation 
fluxes (see e.g. Ref. \cite{Fornasa:2007nr} for a review).
In particular, if the crossing timescale $ t_{cr} $ (the time needed for a 
DM particle to cross the central part of the halo) is much smaller than the 
BH growth timescale $ t_{growth} $, DM particles conserve their angular 
momenta and their radial actions (the integral of the radial velocity over
one closed orbit), and move to orbits which are closer to the BH, thus
enhancing their number density and annihilation rate.

The first study of adiabatic contraction was performed by Young in 1980
\cite{Young:1980} in the context of stellar systems. In subsequent studies 
also DM has been taken into account 
\cite{Quinlan:1994ed,Sigurdsson:2003wu,MacMillan:2002sp} and two different 
situations have usually been considered:
\begin{itemize}
\item ``analytical cores'', in which case the initial DM profile can 
be Taylor-expanded around the BH as: 
$ \rho_\chi(r) \approx \rho_{\chi,0} + 1/2 \rho_{\chi,0}^{\prime\prime} r^2 +
\mathcal{O}(r^3) $,
\item $ \gamma $-models, in which the inner initial DM profile is a power-law,
$ \rho_\chi(r) \propto r^{-\gamma} $.
\end{itemize}

The result of the adiabatic contraction is different in the two cases. A
strong density enhancement, or \emph{spike}, is produced in both cases, but 
the final slope is steeper for $ \gamma $-models. The difference is due to 
the different behavior of the phase-space distribution function $ f(E,t) $. 
In the case of $ \gamma $-models, $ f $ diverges in the limit 
$ E \rightarrow \Phi(0) $, meaning that cold orbits (those orbits populated 
by particles with very low velocities) have a large occupation number. The 
final DM profile for a $ \gamma $-model is a new power-law 
$ \rho_\chi \propto r^{-\gamma_{sp}} $ with a steeper slope. The value of 
$ \gamma_{sp} $ depends on the initial slope as in the following equation:
\begin{equation}
\gamma_{sp}=\frac{9-2\gamma}{4-\gamma}.
\end{equation}

The effect of SMBH growth at the Galactic center on DM density profiles 
is shown in Fig. \ref{fig:spike} for an initial NFW profile ($ \gamma=1 $). 
The final profile in this case is a power-law of index $ \gamma_{sp}=7/3 $ 
within a region of radius $ r_{sp} \sim 0.2 r_h $, where $ r_h $ is the
radius that encloses a DM mass equal to twice the mass of the central BH.

There is an upper limit to the DM density that can be achieved around the 
SMBH (see Fig. \ref{fig:spike}). In fact, the power-law solution is valid 
only down to a radius equal to the last stable orbit around the SMBH and, 
furthermore, DM annihilation itself sets an upper limit on the DM density
of order $ m_\chi/\sigma v(t-t_f) $, so that the density is saturated at a 
cut-off radius of 
\begin{equation}
r_{cut}={\rm Max} \left[ 4 R_{\rm Schw}, r_{\rm lim} \right] ,
\end{equation}

where $ R_{\rm Schw} $ is the Schwarzschild radius of the IMBH 
$ R_{\rm Schw} = 2.95 \,{\rm km} \, (M_{\rm bh}/\msun) $.

Any spike forming at the center of a galactic halo would inevitably be 
affected by dynamical processes that tend to deplete the DM density 
\cite{Ullio:2001fb,Merritt:2002vj,Bertone:2005hw}. For instance, a merger 
between two halos, the interaction with a globular cluster or the presence 
of a bar are all able to disrupt the distribution of the cold particles that 
constitute the spike, efficiently damping the enhancement 
\cite{Ullio:2001fb,Merritt:2002vj}.
Furthermore, the BH may not form exactly at the center of the DM distribution 
\cite{Ullio:2001fb}, in which case the orbits that the BH crosses in its 
spiraling in will be depleted and the final enhancement will be much 
shallower than depicted in Fig. \ref{fig:spike}. Finally, one should also 
take into account the interactions of the DM particles with baryons 
\cite{Merritt:2002vj}: transferring energy to the DM particles, the effect is, 
as before, of heating the DM so that the particles in the spike will be able to
leave the central region. When also including the effect of DM annihilations, 
it is possible to show that any DM overdensity around a BH at the center of a 
galactic halo can hardly survive until today \cite{Bertone:2005hw}, although 
interactions of baryons can actually lead to the re-growth of mild 
overdensities known as DM {\it crests} \cite{Merritt:2006mt}.  
Our approach is to account for some of these mitigating factors by considering 
only those BHs that formed at high redshift and were found never to interact 
with any other BH during the hierarchical merging process, as discussed in 
Ref.~\cite{Bertone:2005xz}. Moreover, at least for scenario B, off-center
formation is not possible for IMBHs and interactions of DM with baryons have
only minor effects, so that, in contrast to SMBHs, IMBHs can effectively
retain their mini-spike.

\begin{figure*}
\begin{center}
\includegraphics[width=0.95\textwidth]{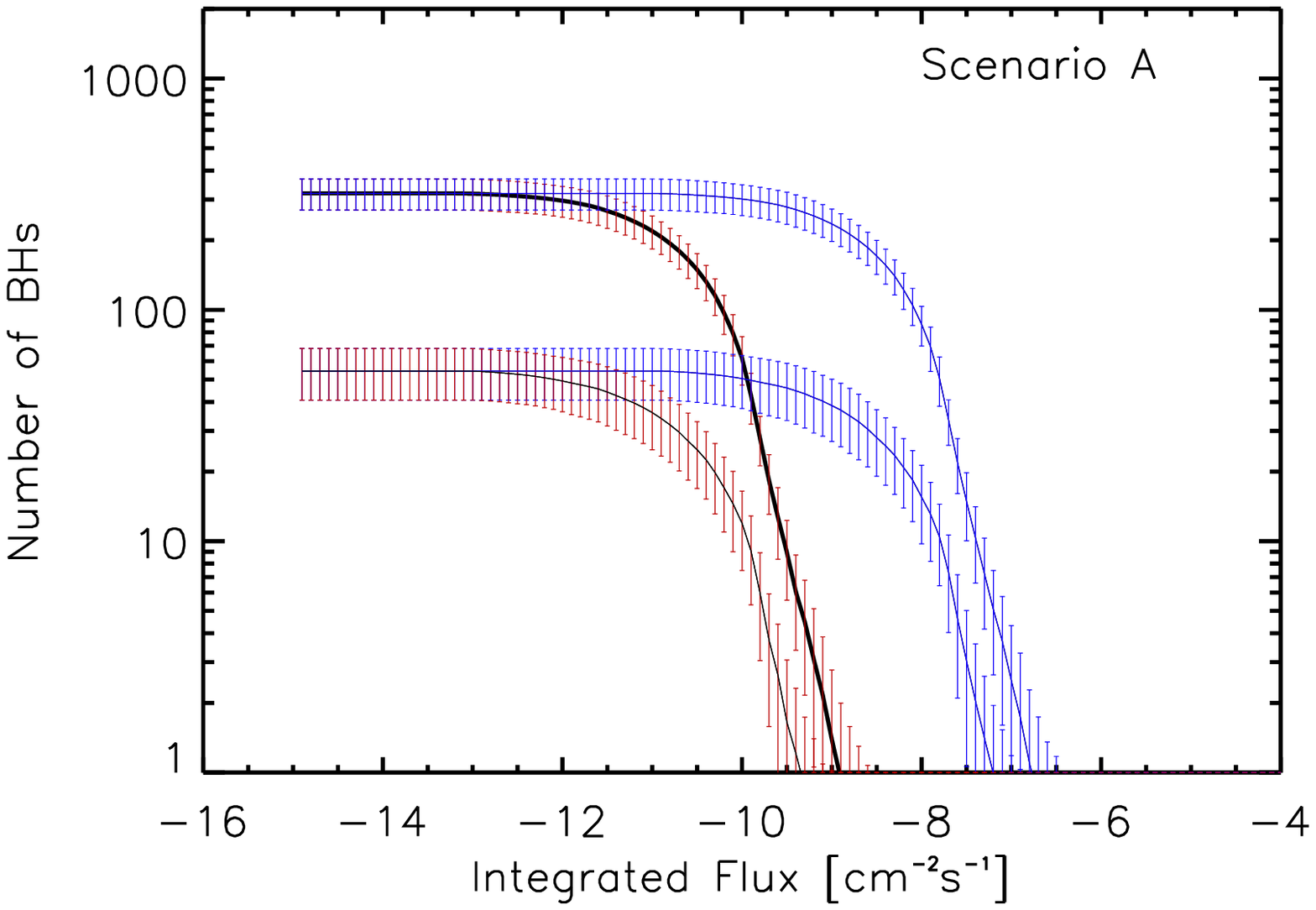}
\caption{\label{fig:detectable_IMBHs_A}  IMBHs integrated luminosity
function, i.e. number of IMBHs producing a gamma-ray flux larger than
a given flux $ \Phi $ (above 1 GeV), as a function of $ \Phi $, for
scenario A. For each curve we also show the $ 1\sigma $ scatter among
the different realization of the Milky Way-size host DM halo. The blue (red)
curve refers to a DM particle with a mass of 100 GeV (1 TeV) and a
cross section of $ \langle \sigma v \rangle = 3 \times 10^{-26}
\mbox{cm}^{-3} \mbox{s}^{-1} $ ($ \langle \sigma v \rangle = 3 \times
10^{-29} \mbox{cm}^{-3} \mbox{s}^{-1} $) . The upper (lower) set of
lines corresponds to a redshift of formation $ z_f=18 $ (15). The
figure can be interpreted as the number of IMBHs detectable by
experiments with a point source sensitivity $ \Phi $ (above 1 GeV) as
a function of $ \Phi $.}
\end{center}
\end{figure*}

\begin{figure*}
\begin{center}
\includegraphics[width=0.95\textwidth]{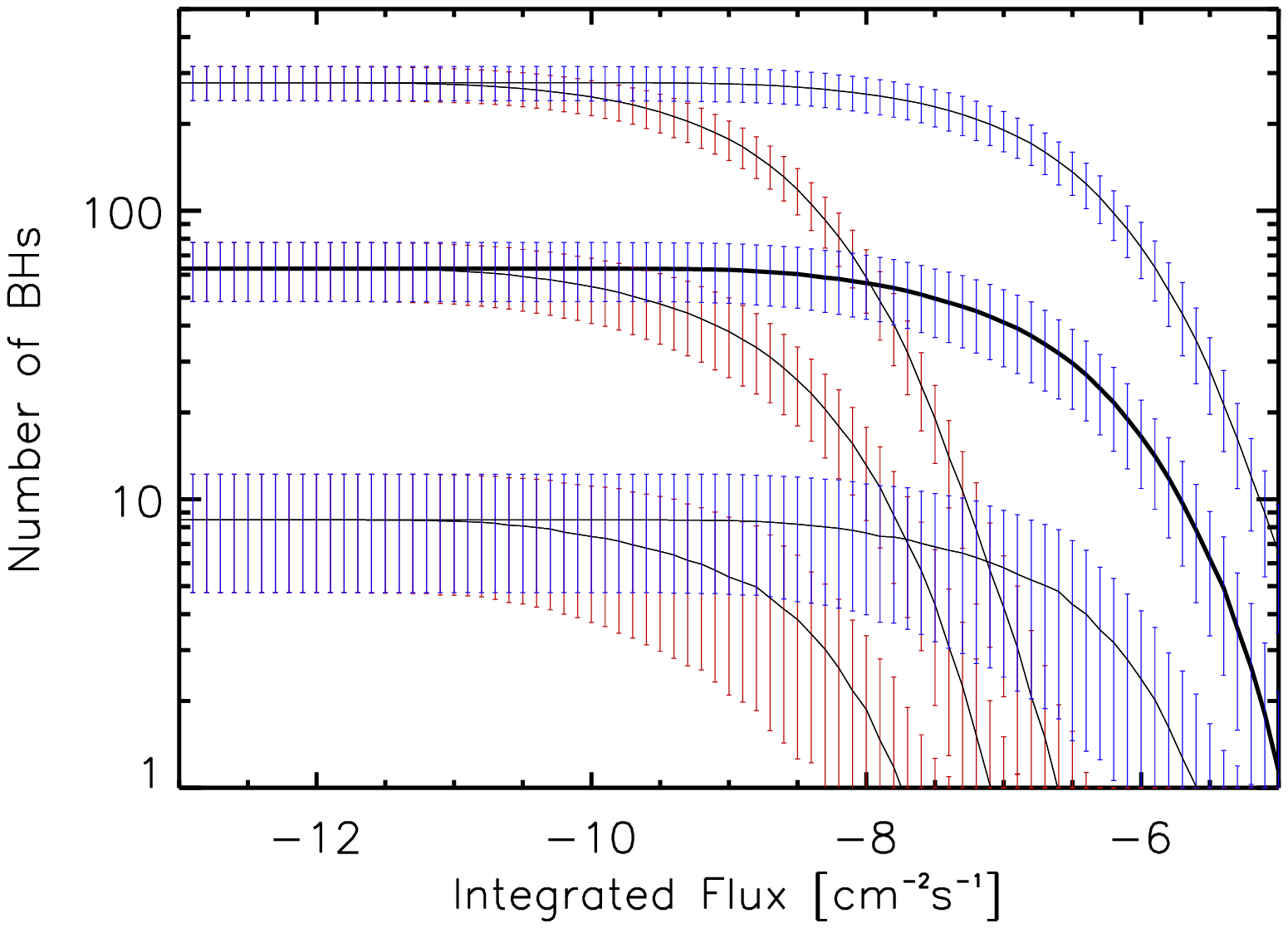}
\caption{\label{fig:detectable_IMBHs}  IMBHs integrated luminosity
function, i.e. number of IMBHs producing a gamma-ray flux larger than
a given flux $ \Phi $ (above 1 GeV), as a function of $ \Phi $, for
scenario B. For each curve we also show the $ 1\sigma $ scatter among
the different realization of the Milky Way-size host DM halo. The blue (red)
curve refers to a DM particle with a mass of 100 GeV (1 TeV) and a
cross section of $ \langle \sigma v \rangle = 3 \times 10^{-26}
\mbox{cm}^{-3} \mbox{s}^{-1} $ ($ \langle \sigma v \rangle = 3 \times
10^{-29} \mbox{cm}^{-3} \mbox{s}^{-1} $) . The middle (upper, lower)
set of lines corresponds to a redshift of reionization $ z_r=12 $ (7,
17). The figure can be interpreted as the number of IMBHs detectable
by experiments with a point source sensitivity $ \Phi $ (above 1 GeV)
as a function of $ \Phi $.}
\end{center}
\end{figure*}

\section{Detection of DM around IMBHs}
\label{sec:Galactic_IMBHs}
The flux of secondary particles produced by DM annihilation around an IMBH can 
be written as
\begin{equation}
\frac{d\Phi}{dE}(E)=\frac{\langle \sigma v \rangle}{8 \pi m_\chi^2}
\frac{dN_\gamma}{dE}(E) \int \rho_{sp}^2(r) dV \, ,
\label{eqn:flux}
\end{equation}

where $ \langle \sigma v \rangle $ is the thermally-averaged annihilation 
cross section, $ m_\chi $ is the mass of the DM particle and $ dN_\gamma/dE $ 
is the number of secondary particles produced per annihilation. 
The integral extends from the \emph{cut radius} $ r_{cut} $ (i.e. the radius 
where the DM density reaches the annihilation plateau in Fig. \ref{fig:spike}) 
to the spike radius $ r_{sp} $.
Inserting the appropriate reference values, Eq.~\ref{eqn:flux} can be recast as
\begin{eqnarray}
\frac{d\Phi}{dE}(E) & = & \Phi_0 \frac{dN_{\gamma}}{dE}
\left( \frac{\langle \sigma v \rangle}{10^{-26} \mbox{cm}^3 \mbox{s}^{-1}}
\right) \left( \frac{m_{\chi}}{100 \mbox{ GeV}} \right)^{-2} \\
& & \left( \frac{d}{\mbox{kpc}} \right)^{-2}
\left( \frac{\rho_{sp}(r_{sp})}{10^2 \mbox{ GeV cm}^{-3}} \right)^2
\left( \frac{r_{sp}}{\mbox{pc}} \right)^{14/3} \nonumber \\
& & \left( \frac{r_{cut}}{10^{-3} \mbox{pc}} \right)^{-5/3}, \nonumber
\end{eqnarray}
with  $ \Phi_0=9 \times 10^{-10} \mbox{cm}^{-2} \mbox{s}^{-1} $, a 
normalization factor that makes it evident that IMBHs can be bright 
DM annihilation sources, possibly as bright as the entire MW.

The fact that the cut radius itself depends on the mass and the cross section
of the DM candidate alters the na\"ive dependence of the annihilation flux from
these quantities, and it is easy to show that in this case 
$ d\Phi/dE \propto \langle \sigma v \rangle^{2/7} m_{\chi}^{-9/7} $. The
weak dependence of the predicted annihilation flux on the particle physics 
parameters is one of the intriguing features of the mini-spikes scenario.

In Figs. \ref{fig:detectable_IMBHs_A} and \ref{fig:detectable_IMBHs} the
number of IMBHs associated with a gamma-ray flux (integrated above 1 GeV)
larger than $ \Phi $ is plotted as a function of $ \Phi $ itself. With
respect to the similar Figs. 4 and 5 in Ref. \cite{Bertone:2005xz}, 
we consider here the most recent cosmological parameters from WMAP5,
and find a significant decrease of the number of IMBHs .
In order to show the effect of changing the cosmological parameters we have
considered two representative values of the redshift of formation $ z_f $
for the scenario A and we have varied the reionization redshift $z_{r}$ for 
the scenario B in the WMAP5 $3\sigma$ range \cite{Komatsu:2008hk}.
The total number of IMBHs for the different choices of astrophysical 
parameters is shown in Table \ref{tab:number}.

\begin{table}[h]
\centering
\begin{tabular}{cccc}
\hline
Scenario & $ z $ & $ N_{BH} $ & $ 1\sigma $ scatter \\
\hline
A & 18 & 319 & 49 \\
A & 15 & 54 &  13\\
B & 7 & 278 &  37 \\
B & 12 & 62 & 14 \\
B & 17 & 8.4 & 3.7 \\
\hline
\end{tabular}
\caption{Number of unmerged IMBHs in a Milky-Way like halo at $ z=0 $.
The four columns indicate the formation scenario, the formation 
(reionization) redshift in the case of scenario A (B), the number
of IMBHs and the $ 1\sigma $scatter among realizations}.
\label{tab:number}
\end{table}

Figs. \ref{fig:detectable_IMBHs_A} and \ref{fig:detectable_IMBHs}
can also be loosely interpreted as the number of objects that can be detected 
with an experiment of sensitivity $ \Phi$. 
For EGRET, $\Phi_{EGRET} \approx 3 \times 10^{-8} \mbox{cm}^{-2} \mbox{s}^{-1} $,
while for Fermi 
$ \Phi_{Fermi} \approx 3 \times 10^{-10} \mbox{cm}^{-2} \mbox{s}^{-1} $.
The fact that the scenario is within the reach of EGRET,  means that some of
the unidentified EGRET sources could be interpreted as IMBHs. If this 
interpretation was confirmed by Fermi, which is expected to see many more 
sources, the identification of a population of point-like sources with the 
same energy spectrum would provide strong evidence for a DM annihilation 
signal. 
We stress that the number of detectable sources depend on the
formation scenario but also on other poorly constrained processes, such as 
the merger history of IMBHs in the galaxy and the initial distribution and 
dynamical evolution of the surrounding DM. Furthermore, from a particle 
physics point of view, we recall that the DM annihilation cross section can 
largely differ from the thermal value in presence of efficient coannihilations 
\cite{Griest:1990kh}, significant Sommerfeld enhancements 
\cite{Sommerfeld,Hisano:2003ec,Cirelli:2007xd,Lattanzi:2008qa}, in cases where 
DM particles are not thermal relics, or in non-standard cosmological 
scenarios \cite{Kamionkowski:1990ni,Catena:2004ba,Gelmini:2006pq}. 

Even in the conservative case where none of the EGRET sources 
is associated with a mini-spike, there are models where the Fermi satellite 
may detect this class of objects, do to the steep decline of the luminosity 
function in Figs. \ref{fig:detectable_IMBHs_A} and \ref{fig:detectable_IMBHs}.

The first experimental constraints on the mini-spikes scenario have been
obtained through a survey of the inner Galactic plane at photon energies above 
100 GeV performed by the H.E.S.S. array of Cherenkov telescopes from 2004 
to 2007 \cite{Aharonian:2008wt}. About 400 hours of data have been 
accumulated in the region between -30 and +60 degrees in Galactic longitude, 
and between -3 and +3 degrees in Galactic latitude, and a H.E.S.S. sensitivity 
map was computed for DM annihilations. The data exclude scenario B at a 
90\% confidence level for DM particles with velocity-weighted annihilation 
cross section $ \sigma v \gtrsim 10^{28} $ cm$ ^3 $ s$ ^{-1} $ and mass 
between 800 GeV and 10 TeV.

\begin{figure}
\begin{center}
\includegraphics[width=0.5\textwidth]{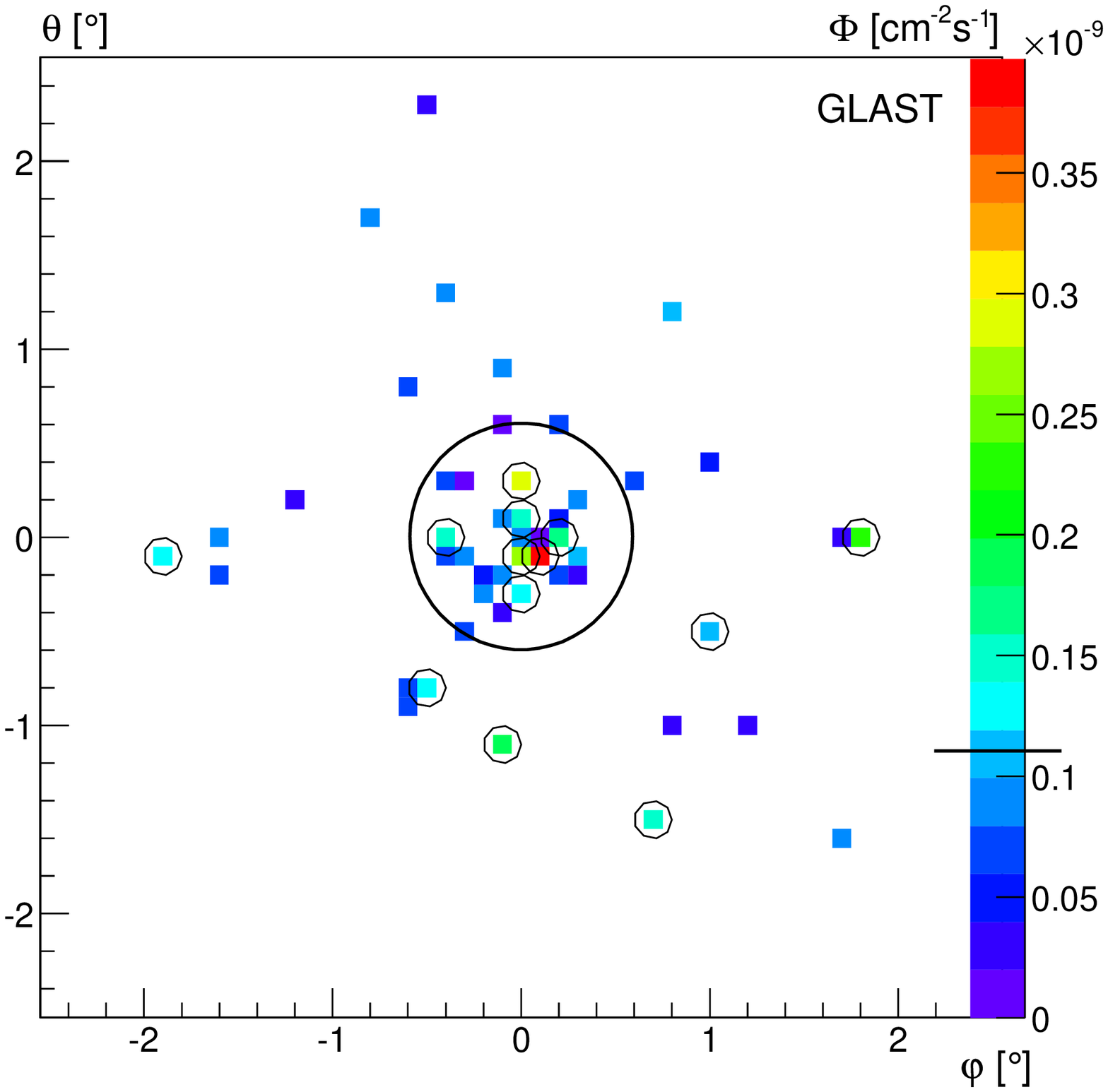}
\caption{\label{fig:IMBHs_Andromeda} Map of the gamma-ray flux in
units of photons $\mbox{cm}^{-2}\mbox{s}^{-1} $, from DM annihilations
around IMBHs in the Andromeda galaxy M31, relative to one random
realization of IMBHs in M31. The size of the bins is $0.1^{\circ}$ and
the threshold is 4 GeV as appropriate the best angular resolution of
Fermi (still labeled as GLAST in the figure). The circles
highlight IMBHs within the reach of Fermi for a 5$\sigma$
detection in 2 months. The big circle shows for comparison the M31
scale radius $ r_s $.}
\end{center}
\end{figure}

As for the prospects for detecting mini-spikes with the Fermi telescope,
a dedicated analysis of the minimum detectable annihilation flux has been
presented in Ref. \cite{Bertone:2006kr} and further discussed in the 
pre-launch estimates of the Fermi sensitivity to DM annihilation signals
\cite{Baltz:2008wd}.  The mock IMBH catalogs obtained for the Milky Way have 
been adapted, by a suitable rescaling, to the population of IMBHs hosted by
the Andromeda galaxy \cite{Fornasa:2007ap}, leading to the prediction that
the M31 should host about 65 IMBHs (again this estimate strongly depends on 
the assumed cosmological parameters). Detectability toward Andromeda is 
reduced greatly due to the comparably large distance to the Andromeda galaxy.  
On the other hand, in the case of detecting a handful of such objects, 
they will all be located within $ \sim 3^\circ $ from the center of Andromeda, 
which may be an interesting spatial signature (see 
Fig. \ref{fig:IMBHs_Andromeda}).

\subsection{Neutrinos from IMBHs}
DM annihilation can also produce neutrinos, both directly or through the
decay of secondary particles. Being strong annihilation "boosters", IMBHs may 
also be bright neutrino sources \cite{Bertone:2006nq,Bertone:2007ei}, which 
can be detected by high energy neutrino telescopes such as ANTARES and IceCube 
or the future Km3.  Neutrino telescopes detect neutrinos through the 
observation of muons produced by charged current interactions of neutrinos 
with the nuclei around the detector. Accounting for oscillations we can write 
the muon flux as:
\begin{eqnarray}
\frac{d\Phi_{\nu_\mu}}{dE}(E) & = & \frac{\langle \sigma v \rangle}{8 \pi
m_\chi^2} \int \rho^2(r) dV \\
& & \sum_{\ell \in \{e,\mu,\tau\}} \mathcal{P}(v_\ell \rightarrow \nu_\mu)
\frac{dN_{\nu_\ell}}{dE}(E), \nonumber
\end{eqnarray}

where $ \mathcal{P}(v_\ell \rightarrow \nu_\mu) $ is the probability of having a
muon neutrino at the detector when at the source a neutrino of the $ \ell $
family is produced, and it can be written in terms of the oscillation
matrix $ U $:
$ \mathcal{P}(v_\ell \rightarrow \nu_\mu)=\sum_{j \in \{e,\nu,\tau\}} |U_{\ell j}|^2 |U_{\mu j}|^2 $.

Finally, the rate of events in a neutrino telescope is
\begin{equation}
R=\int_{E_\mu^{thr}}^{m_\chi} dE_\nu \int_0^{y_\nu} dy A(E_\mu) P_\mu(E_\nu,y)
\frac{dN_\nu}{dE_\nu}(E_\nu),
\end{equation}

where $ y_\nu=1-E_\mu^{thr}/E_\nu $, $ E_\mu^{thr} $ is the experimental
threshold for muons, $ A(E_\mu) $ is the effective area and $ P_\mu(E_\nu,y) $
is the probability that a muon neutrino interacts with a nucleon producing a
muon of energy $ E_\mu=(1-y)E_\nu $ (see Ref. \cite{Bertone:2006nq} for
details). Results are summarized in Fig. 2 of Ref. \cite{Bertone:2006nq},
similar to Fig. \ref{fig:detectable_IMBHs} in the case of gamma-rays.
Those results can be easily extended to the new models presented here, 
the main consequence being a smaller number of detectable sources.

\section{Contribution of IMBHs to the EGB and their angular spectrum}
\label{sec:Angular_spectrum}

\begin{figure*}
\begin{center}
\includegraphics[width=0.45\textwidth]{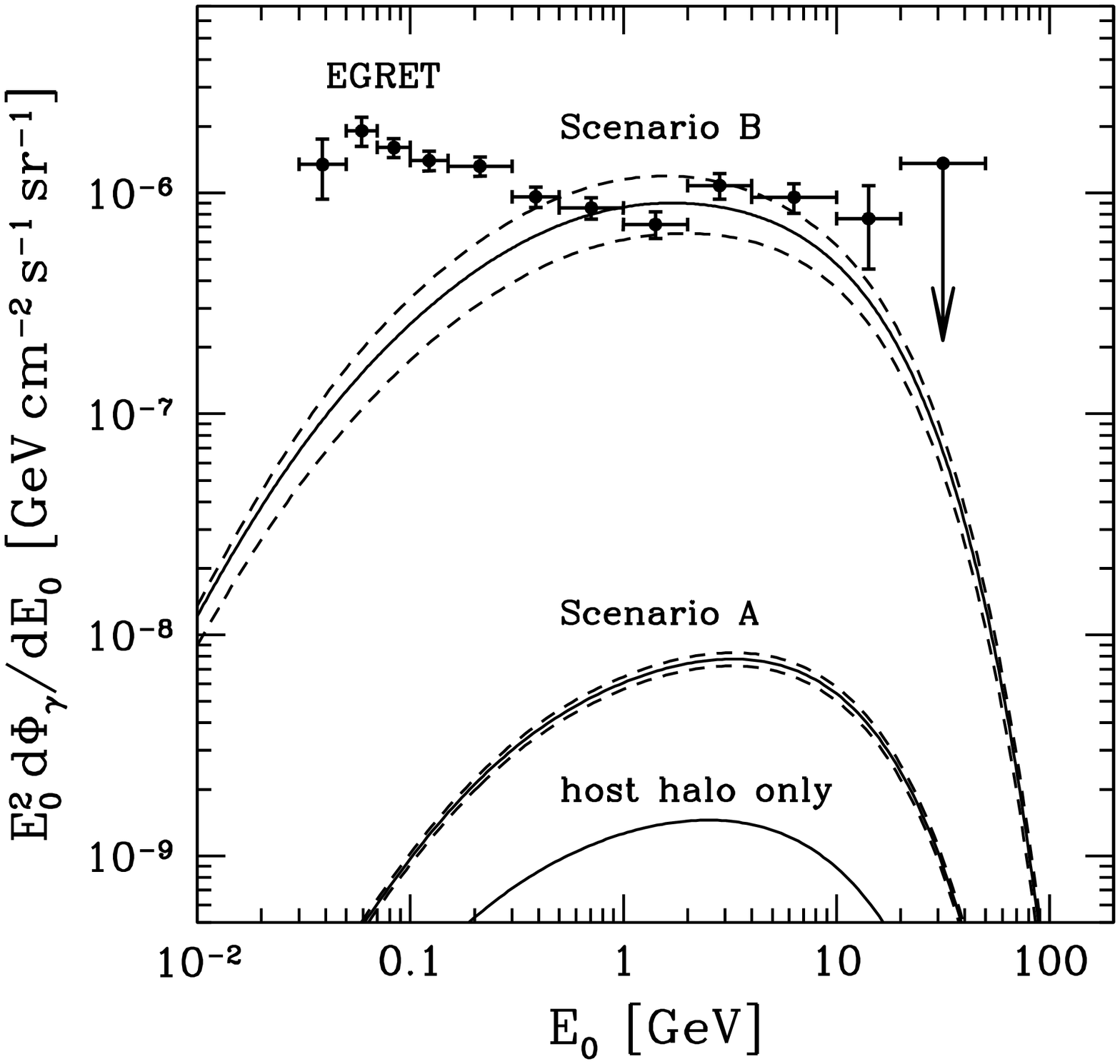}
\includegraphics[width=0.45\textwidth]{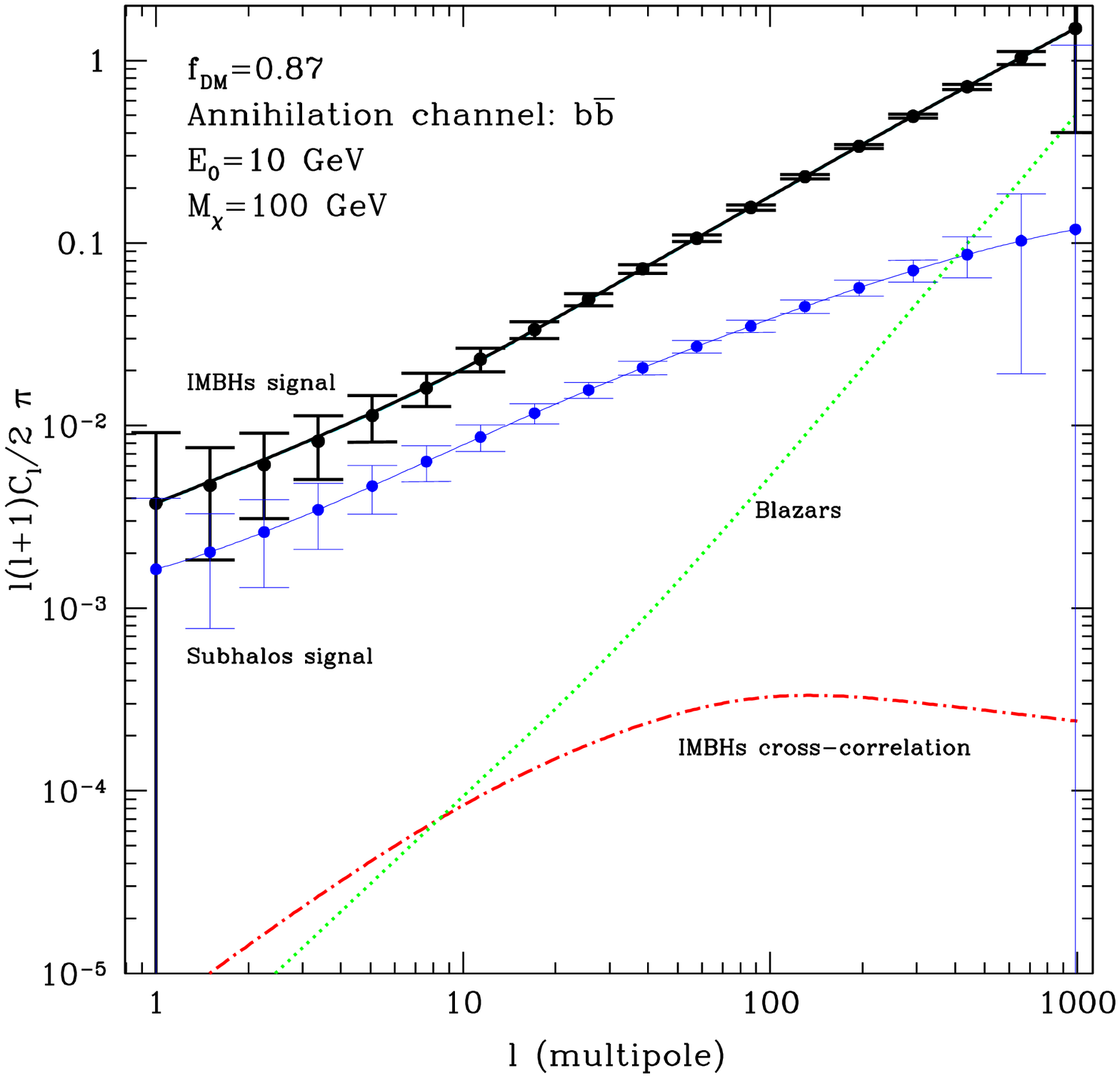}
\caption{\label{fig:CGB}  {\it Left:} Contribution to the EGB from DM
annihilations around IMBHs for scenario A and B. $ m_\chi=100 \mbox{
GeV} $, $ \langle \sigma v \rangle = 3 \times 10^{-26} \mbox{cm}^3
\mbox{s}^{-1} $ and DM annihilations into $ b\bar{b} $ have been
considered. Taken from \cite{Horiuchi:2006de}. {\it Right:} Angular
power spectrum of the EGB from DM annihilations around IMBHs at an
energy $ E_0=10 $ GeV. The dashed line shows the DM contribution, the
dotted one is for unresolved blazars and the dot-dashed is the
cross-correlation term. The black solid line is the total signal and
the error bars are for 2-years of Fermi data. The solid blue is
the signal for DM annihilations in subhalos. Taken from
\cite{Taoso:2008qz}.}
\label{fig:Anisotropies CGB} 
\end{center}
\end{figure*}

As we have seen in the previous sections, IMBHs are generically predicted in 
scenarios that seek to explain the existence of SMBHs at the centers of 
galaxies. It is therefore possible, in principle, that the gamma-ray flux 
from DM annihilations around IMBHs in all DM halos, and at all redshift, add 
together and contribute significantly to the diffuse Extra-galactic Gamma-ray 
Background (EGB) discovered by EGRET \cite{Sreekumar:1997un}.

The origin of this gamma-ray emission is actually unknown, and although
unresolved blazars are often considered as the most likely sources of the 
EGB, the most recent analysis suggest that they can not account for more than 
25-50\% of the measured EGB \cite{Narumoto:2006qg}. Besides standard 
astrophysical sources, like galaxies or clusters of galaxies 
\cite{Liang:2002fj,Rephaeli:2008jp,Pavlidou:2002va}, DM annihilations have
been extensively studied as possible candidates to explain the EGB emission
\cite{Ullio:2001fb,Bergstrom:2001jj,Taylor:2002zd,Fornasa:2009qh}.

The contribution from cosmological DM halos to the EGB is constrained
to be rather low, in order to satisfy the observational bounds at the center
of our Galaxy \cite{Ando:2005hr}, even when the effect of spikes around SMBHs
is taken into account. The presence of cosmological DM clumps could enhance 
the signal but according to the most recent results from DM simulations the 
expected boost factors are small (see 
Refs. \cite{Diemand:2008in,Springel:2008by} and also the results in
Ref. \cite{Fornasa:2009qh}). A sizeable diffuse emission is instead produced
by the population of unresolved DM clumps hosted in our Galaxy
\cite{Fornasa:2009qh,SiegalGaskins:2008ge}.

As for IMBHs, they can account for a large fraction of the measured EGB 
\cite{Horiuchi:2006de}. The mean intensity as a function of the energy, 
$ \langle I(E)_{DM} \rangle $ produced by DM annihilations in cosmological 
IMBHs is given by:
\begin{equation}
\langle I(E)_{DM}\rangle=\int \frac{c dz}{H(z)}
W\left(E[1+z],z\right),
\label{eqn:Int}
\end{equation}

where
\begin{equation}
W(E,z)=\frac{\left(\sigma v\right)}{8\pi m_{\chi}^2}
\frac{dN_{\gamma}}{dE}\left( E[1+z]\right) e^{-\tau(E[1+z],z)}
\Delta^2(z).
\label{eqn:Intensity}
\end{equation}

and the optical depth $ \tau $ parametrizes the gamma-ray attenuation due to
interactions with the extra-galactic background light (see e.g.
Ref. \cite{Salamon:1997ac}).

The factor $ \Delta^2(z) $ in Eq. \ref{eqn:Intensity} contains the information
on the DM spike profile $ \rho_{sp} $ around each IMBH and the comoving number 
density $ n(z) $:
\begin{equation}
\Delta^2(z)=n(z) \int_{r_{cut}}^{r_{sp}} \rho_{sp}^2 (r) d^3 r.
\end{equation}

The latter input cannot be computed in a straightforward manner using 
our current simulation techniques, but it can be modeled in an approximate 
manner based upon their results. For example, the BH formation models assume 
that initial BH formation is peaked at a certain redshift, $ z_f $, to a good 
approximation, and is effective only in DM halos with masses above a certain 
threshold $ M_{thr} $. These two parameters depend upon the specific formation 
scenario. Following this prescription and employing the halo mass function
$ dn/dM(M,z) $ predicted from extended Press-Schechter theory
\cite{Sheth:1999mn}, the comoving number density at the formation redshift is:
\begin{equation}
n(z_{f})=\int_{M_{thr}}^{\infty} dM \frac{dn}{dM}(M,z=z_{f}),
\end{equation}

where it is conservatively assumed that only one IMBH is formed per halo.

The number of IMBHs in present halos depend on the formation history of the
host halo and is affected by the occurrence of BHs mergers. In good
approximation it follows a linear dependence on the mass of the host halo,
with a normalization $ N_{bh} $ for the Milky Way that is obtained from 
simulations.
The present IMBHs comoving number density is thus:
\begin{equation}
n(0)=\int_{M_{min}}^{\infty} dM \frac{dn}{dM}(M,z=0) N_{bh}
\frac{M}{M_{MW}}.
\label{eqn:nz0}
\end{equation}

For a generic redshift $ z $ the IMBHs comoving number density can be
parameterized assuming a power-law behavior which interpolates the BH number
density at the formation redshift and at $ z=0 $:
$n(z)=n(z_f)\left[\left(1+z\right)/\left(1+z_f\right) \right]^{\beta}.$

With this recipe and based on the results of the simulation of
Ref. \cite{Bertone:2005xz} the authors in Ref. \cite{Horiuchi:2006de} computed 
the contribution of IMBHs to the EGB for the two IMBHs formation scenarios 
presented in Sec. \ref{sec:Formation_scenarios} and for different choices of 
DM particle physics properties. In Fig. \ref{fig:CGB}, we show that in the 
conservative case of scenario A the expected signal is largely below the EGB 
measurements. However, for scenario B the predicted diffuse emission is at 
the level of EGRET observations already for a standard particle physics 
setup, e.g. assuming a "thermal" annihilations cross section
$\langle \sigma v \rangle = 3 \times 10^{-26} \mbox{cm}^{3} \mbox{s}^{-1} $
and $ m_\chi=100 \mbox{ GeV} $.

As with the number of Galactic IMBHs, these predictions are subject to 
relatively large astrophysical uncertainties. Nonetheless, it is remarkable 
that models with astrophysical and particle physics parameters that have been 
fixed for independent reasons, the DM signal from cosmological IMBHs is 
comparable with the EGB flux extracted from EGRET observations. Of course, in 
order to detect effectively DM annihilations in the EGB it is necessary to 
distinguish the DM signal from other astrophysical emissions, such as that 
originated by blazars. Unfortunately, this is a difficult program to pursue 
using only spectral information, unless the DM signal is accompanied by 
striking spectral features, like those induced by gamma-ray line emission 
(see Ref. \cite{Horiuchi:2006de} for computations in IMBHs scenarios).

This problem can however be overcome by studying the angular correlations of 
the diffuse extra-galactic emission, as in Ref. \cite{Ando:2005xg}.  
In fact, the angular power spectrum of anisotropies from DM annihilations and 
from blazars are different, due to their different spatial distribution, 
energy spectra and radial emissivity profiles. Later on, other papers computed 
the angular spectrum of the EGB in the case that a substantial amount of the 
average EGB flux is accounted by DM annihilations in extra-galactic halos and 
subhalos \cite{Ando:2006cr} or in Galactic DM clumps 
\cite{SiegalGaskins:2008ge}, or considering both the contributions 
\cite{Fornasa:2009qh}. In all cases, if DM plays a role in the EGB, the 
analysis of the angular power spectrum of gamma-ray anisotropies can nicely 
detect its contribution and with the soon-available full-sky data by 
Fermi this approach will be testable.

In the case of IMBHs, the authors of Ref. \cite{Taoso:2008qz} focused on
scenario B and computed the angular power spectrum induced by DM annihilations
around IMBHs. We refer to the original reference for details on the
computations. As shown in Fig. \ref{fig:Anisotropies CGB} the shape of the
angular power spectrum should help in distinguishing the DM signal from the 
blazar contribution to the EGB. These results refer to an energy of 
observations of $ E_0=10 $ GeV, which is a good compromise between 
maximization of the photon count and minimization of the Galactic foreground,
which tends to masquerade the extra-galactic emission. In addition, natural
DM candidates of around $ 100 \mbox{ GeV} $ produce the largest gamma-ray
yield at these energies. In order to assess more precisely the prospects for
DM detection in the EGB they considered the unresolved blazar contribution as
a known background, motivated by the fact that it should be modeled quite
accurately from the future Fermi catalog of detected blazars. In addition,
the power spectrum from unresolved blazars should be energy independent, due
to the power-law energy dependence of their energy spectra and therefore it
could be calibrated at low energies, where the contribution of DM
annihilations is negligible and then subtracted from the total anisotropy
data. After modeling the blazars contribution to the EGB, if it is assumed
that the remaining fraction $ f_{DM} $ of the EGB flux at the energy of
observation $ E_0 $ is due to DM annihilations, the signal $ C_l^s $ depends
on the DM power spectrum $ C_l^{DM} $ and on the cross correlation between
the DM and blazars signals $ C_l^{Cr} $:
\begin{equation}
C_l^s = f_{DM}^2C_l^{DM} + 2f_{DM}(1-f_{DM})C_l^{Cr}.
\end{equation}

Fig. \ref{fig:Anisotropies CGB} shows the signal as well as the projected 
$ 1\sigma $ error bars after two years of Fermi observations. The 
prospects for distinguishing the signal of DM annihilation around IMBHs in 
the angular EGB power spectrum are encouraging. This conclusion holds for 
different choices of DM mass and annihilation channel and energy of 
detection, provided that DM annihilations substantially contribute to the EGB 
at $ E_0 $, i.e. for $ f_{DM}\gtrsim 0.3 $. Moreover, if the DM signal is 
dominated by DM emission in cosmological subhalos instead that around IMBHs, 
the predicted angular power spectrum is different and for sizeable values of 
$ f_{DM} $ the two scenarios should be distinguishable by Fermi observations.

\section{Discussion and conclusions}
\label{sec:Discussion_conclusions}

\begin{figure}
\begin{center}
\includegraphics[width=0.5\textwidth]{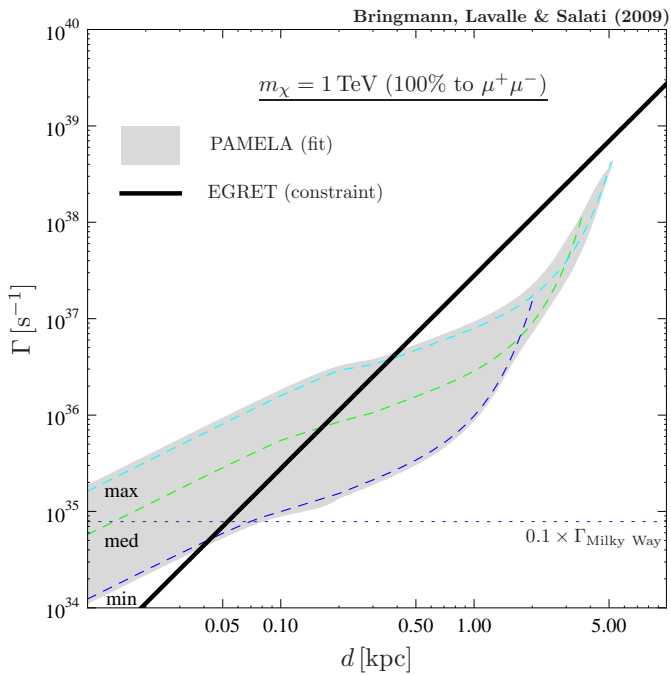}
\caption{\label{fig:Pamela} The solid line show the constraint in the
DM annihilation rate $ \Gamma=1/2 \langle \sigma v\rangle/m_\chi^2
\int d{\bf x} \rho_{sp}({\bf x})$ of a nearby DM overdensity at a
distance $ d $ from the Earth. The shaded area shows the combinations
of $ \Gamma-d $ needed to explain the PAMELA positron excess. The
dashed lines refers to a set of propagation parameters defined in
Ref. \cite{Delahaye:2008ua}. For comparison the dotted line indicates
the 10\% of $ \Gamma $ for the whole MW assuming $ \langle \sigma v
\rangle=3 \times 10^{-26} \mbox{cm}^{-3} \mbox{s}^{-1} $. Taken from
\cite{Bringmann:2009ip}.}
\end{center}
\end{figure}

Indirect DM searches have received a lot of attention recently, 
in large part due to the recent results of experimental collaborations 
such as Fermi, PAMELA, HESS and others. However, unless unambiguous 
spectral features are detected \cite{spectralfeatures}, it will be difficult to 
discriminate any incident flux of high-energy particles from astrophysical 
sources (either known or unknown) and identify a signal due to DM 
annihilation.  The mini-spikes scenario discussed here
may provide an interesting solution to this problem.

In fact, in the previous sections we have shown that IMBHs are promising
targets for DM indirect detection, as a consequence of the large DM
overdensities produced around them by gravitational processes during the
formation of the BH. The properties of these objects and the distribution 
of DM about these objects both depend on the unknown BH formation mechanism.
In Ref. \cite{Bertone:2005xz} the authors investigated in detail two
astrophysical models proposed in the contemporary literature.  This study 
utilized numerical simulations to make predictions for the population of 
IMBHs that should be present in our Galaxy today.  
These objects could appear as bright gamma-ray emitters as a consequence of
the DM annihilations occurring in the dense DM spikes.  Indeed the
prospects for detecting such objects with Fermi and Air Cherenkov Telescopes 
are encouraging. A spectacular signature for a DM signal may come from the
detection of a class of gamma-ray point sources that are not associated with 
astrophysical sources and that share the same energy spectrum.
IMBHs may also be detected through neutrinos or they can produce a sizable
excess in antimatter cosmic-ray fluxes, as studied in
Ref. \cite{Bringmann:2006im}. Moreover, analogous populations of IMBHs
should be present in other galaxies, such as the nearby Andromeda
galaxy and several gamma-ray sources could be detectable.

In general, the gamma-ray fluxes from IMBHs in all DM halos at all redshifts
produce a diffuse gamma-ray background that for the most optimistic scenario
B can explain a large fraction of the observed EGB. This opens the
interesting possibility to detect DM annihilations in the extra-galactic
background. The best strategy would be to analyze not only the EGB energy
spectrum but also the angular correlations of the emission.
In fact, the EGB angular power spectrum could provide a robust method to
distinguish the DM signal from conventional astrophysical emission.

In Ref. \cite{Bringmann:2009ip}, the authors derived constraints 
on the mini-spike scenario from a comparison of predictions based on BZS
models with EGRET data.  These authors claim that they rule out the 
possibility to detect the products of DM annihilation in mini-spikes. Despite 
the bold claim, their analysis only rules out a specific combination of 
cosmological and astrophysical parameters, for only one (out of many) possible 
formation scenarios.  Actually, we showed above that by simply adopting the 
WMAP 5-year cosmological parameters, the expected number of objects 
went down by a factor of 2, and that changing the redshift of formation in 
scenario A, and of ionization in scenario B, substantially modifies the 
predicted number of sources. Moreover, Figs. \ref{fig:detectable_IMBHs_A} and
\ref{fig:detectable_IMBHs} show that one can obtain quite optimistic prospects
for Fermi (with up to $ \sim 200 $ sources in reach of detection) even 
assuming no detection by EGRET.

Still, the constraints of Ref. \cite{Bringmann:2009ip} are extremely useful.  
In particular, these results bear on the possible implications of 
the mini-spikes scenario for the recent measurements of the positron
and electron flux at high energy \cite{Adriani:2008zr,Chang:2008zzr}. 
The possibility that mini-spikes boost the annihilation signal thus 
providing the appropriate normalization of the positron flux is found to be 
viable, as shown in Fig. \ref{fig:Pamela}, and it would be interesting to see 
whether such a result holds for updated mini-spikes models. This would 
allow to circumvent the multi-messenger constraints 
(see, e.g., Refs. \cite{Bertone:2008xr,Bergstrom:2008ag}). 

In conclusion, the mini-spike scenario provides the opportunity to 
discover a new population of sources (IMBHs) and DM annihilations at the same 
time. The most prominent predicted signature of this scenario is unmistakable: 
the observation of many point-like sources, not correlated with the Galactic 
disk, with an identical spectrum. It is therefore important, despite 
the various constraints and the many astrophysical uncertainties,
to search for this population of objects in present and upcoming data,
especially those of the gamma-ray satellite Fermi.

\end{document}